\documentclass[aps,epsfig,nofootinbib,a4paper,12pt,floatfix]{revtex4}
\usepackage[utf8]{inputenc}

\usepackage{hyperref}
\usepackage{setspace}
\usepackage{float}
\usepackage{slashed}
\usepackage{amsmath,esint}
\usepackage{bm}
\usepackage{epstopdf}
\usepackage{latexsym}
\usepackage{amssymb}
\usepackage[utf8]{inputenc}
\usepackage{lmodern}
\usepackage{blindtext}
\usepackage[T1]{fontenc}
\usepackage[dvipsnames]{xcolor}
\usepackage[utf8]{inputenc}
\usepackage[english]{babel}
\usepackage[letterpaper]{geometry}
\usepackage{graphicx}
\usepackage{clipboard}
\newclipboard{myclipboard}
\usepackage{lipsum}
\usepackage[english]{babel}
\usepackage[center]{subfigure}
\usepackage{verbatim}
\usepackage{amsfonts}
\usepackage[symbol]{footmisc}
\def\correspondingauthor{\footnote{Corresponding author.}}
\begin{document}

\title{Scalar Curvature Invariants in Classical and Quantum Gravity }
\author{B. Shakerin$^{a,b}$}
\email{bahram.shakerin@gmail.com}
\author{D.D. McNutt$^{c}$}
\email{david.d.mcnutt@uis.no}
\author{B. Mattingly$^{a,b}$}
\email{brandonmattingly@protonmail.com}
\author{A. Kar$^{a,b}$}
\email{Abinash\_Kar@baylor.edu}
\author{W. Julius$^{a,b}$}
\email{William\_Julius1@baylor.edu}
\author{M. Gorban$^{a,b}$}
\email{Matthew\_Gorban1@baylor.edu}
\author{C. Watson$^{a,b}$}
\email{Cooper\_Watson@baylor.edu}
\author{P. Brown$^{a,b}$}
\email{pbrown3997@cox.net}
\author{J. S. Lee$^{a,b}$}
\email{Jeff\_Lee@baylor.edu}
\author{E. W. Davis$^{b}$}
\email{Eric\_W\_Davis@baylor.edu}
\author{G.~B.~Cleaver$^{a,b}$\correspondingauthor{}}
\email{Gerald\_Cleaver@baylor.edu}

\affiliation{$^{a}$ Physics Department, Baylor
University, Waco, TX 76798-7316, USA}

\affiliation{$^{b}$ Early Universe, Cosmology and Strings \textnormal{(EUCOS)} Group, Center for Astrophysics, Space Physics and Engineering Research \textnormal{(CASPER)}, Baylor University, Waco, TX 76798, USA}
\affiliation{$^{c}$Faculty of Science and Technology,\\
University of Stavanger, 
N-4036 Stavanger, Norway}

\date{\today}
\begin{abstract}
A short review of scalar curvature invariants in gravity theories is presented. We introduce how these invariants are constructed and discuss the minimal number of invariants required for a given spacetime.  We then discuss applications of these invariants and focus on three topics that are of particular interest in modern gravity theories.

\end{abstract}

\maketitle
\section*{\emph{Essay written for the Gravity Research Foundation 2021 Awards for Essays on Gravitation.}}
\thispagestyle{plain}
\pagestyle{plain}
\newpage

In General Relativity (GR), there has been a focus on scalar curvature invariants (hereafter called curvature invariants) which are scalars constructed from the Riemann, Weyl and Ricci tensors and their covariant derivatives \cite{Stephani}. One can find extensive studies on the mathematical structure of these scalars and the way they can be constructed \cite{Stephani,Ref1,Ref2,Cherubini:2003nj,Coley:2009eb}, as well as case studies for different spacetimes \cite{Mattingly1,Mattingly2,Mattingly3,Mattingly4,Overduin,Ghodrati:2020mtx,Chrusciel:2019xuf,Henry:1999rm,Aksteiner:2018vze,Abdelqader:2014vaa,Page:2015aia,McNutt:2017gjg,Coley:2017vxb,Tavlayan,McNutt:2017paq}. As there are an infinite number of invariants that can be generated, one may ask: what is the minimum number of algebraically independent invariants required in order to generate every other invariant for any given metric? Such a finite set is called a complete set.  

In the case of algebraic curvature invariants, which are built from only the Ricci tensor and Weyl tensor, it was conjectured that at most 14 independent scalars are needed to cover all possible sub-spaces \cite{P1,P2,P3,P4,P5,P6}. These independent scalars were categorized into four different types: the Ricci scalar, two complex invariants $I$ and $J$ formed from the Weyl spinor, and three real invariants $I_6$, $I_7$ and $I_8$ formed from the trace-free Ricci spinor and three complex invariants $K$, $L$ and $M$. Carminati-McLenaghan (CM) investigated the aforementioned proposal and showed that the 14 previously suggested invariants fail to describe solutions admitting a perfect fluid \cite{Ref1}. Thus, CM came up with 16 necessary independent invariants to cover all sub-spaces including the degenerate cases such as the Einstein-Maxwell system and the perfect fluid. Later, Zakhary-McIntosh (ZM) \cite{Ref2} constructed  the first complete set of algebraic invariants for all possible type of metrics. The tensorial and spinorial forms of ZM invariants are given in Refs. \cite{Overduin,Eichhorn2} and \cite{Ref2}, respectively. Less is known about the higher order curvature invariants, which involve the covariant derivatives of the Ricci tensor and Weyl tensor up to a particular order. Most efforts here have focused on ``brute-force'' calculations to determine a complete set of invariants up to a particular order \cite{martin2008invar}.

Curvature invariants can give insight into the physical properties of solutions in GR and we will provide here a brief, yet incomplete, list of their application in gravity theories. The CM invariants have been used to study the behaviour of the Riemann tensor for Lorentzian traversable wormholes and warp drives \cite{Mattingly1,Mattingly2,Mattingly3,Mattingly4}, and the ZM invariants have been used to study charged and rotating black holes \cite{Overduin}. Curvature invariants have also been shown to detect the horizon for all stationary black holes and a large class of dynamical black holes \cite{Abdelqader:2014vaa,Page:2015aia,McNutt:2017gjg,Coley:2017vxb,Tavlayan,McNutt:2017paq} and distinguish this hypersurface from a wormhole throat \cite{McNutt}.

In this essay, we will discuss the role of curvature invariants in three important aspects of gravitational theories. We will discuss how curvature invariants can be used to study the nature of singularities \cite{Mattingly1,Mattingly2,Mattingly3,Mattingly4,Overduin,Ghodrati:2020mtx,Chrusciel:2019xuf,Henry:1999rm,Aksteiner:2018vze}, their relationship to gravitational entropy of a spacetime through the Weyl curvature conjecture \cite{Gron1,Gron2,Gron3,Romero,Li,shakerin}, and their use in quantum gravity in dynamical singularity-resolution \cite{Eichhorn,Calmet:2020vuh,Falls} and in loop quantum cosmology \cite{Singh}.

It has been proven that a large class of spacetime metrics necessarily must admit singularities \cite{Hawking1,Hawking2,Senovilla:2006db}. In GR, when a metric has a singularity, it can be very difficult to determine whether this is a physical singularity or merely an artifact of the coordinate system one chooses. The latter are known as coordinate singularities and they can be removed using a coordinate transformation. In the case of the Schwarzschild metric, after rewriting the Schwarzschild solution in terms of the Kruskal-Szekeres coordinates \cite{Kruskal,Szekeres}, it becomes clear that $r = 2M$ (for $c = G =1$) is a coordinate singularity, not a physical one. In most recent GR textbooks, the Kretschmann scalar, $R^{\alpha\beta\mu\nu}R_{\alpha\beta\mu\nu} = \frac{48M^2}{r^6}$, is introduced to deal with the problem. Since this is not infinite at the event horizon of $r = 2M$, we can easily conclude that $r = 2M$ is just a coordinate
singularity and it can be removed via a proper coordinate transformation. Similarly, the singularity at $r = 0$ is present in the Kretschmann scalar and must be a real singularity. But the question is still open, what about the cases where finding a proper transformation is challenging to rewrite the metric in a conformally equivalent way to the initial metric and studying the resulting Penrose diagrams? Is there
any way one could study the nature of the underlying metric singularities? A solution is to work with quantities that are independent of the choice of coordinates and here the curvature invariants are ideally suited to this task. 

In two long technical essays, Penrose \cite{Penrose1,Penrose2} raised some concerns about a conceptual and fundamental conflict between the second law of thermodynamics and cosmology. In an attempt to summarize Penrose's points, P.C.W. Davies \cite{Davies} sent a letter to Nature to address the issue. According to Davies, in the big-bang model, material content of the universe began in a thermodynamic equilibrium. The universe today is highly ordered and far from equilibrium. In other words, a gravity-dominated evolution may violate the second law of thermodynamics as long as only thermodynamic entropy is taken into account. Penrose suggested that this problem could be solved by assigning entropy to the gravitational field itself. Penrose also conjectured that  the Weyl curvature tensor could be used as a measure of the gravitational entropy. One can express this conjecture in terms of the ratio of Weyl and Ricci curvature invariants \cite{Wainwright}, $P^2=\frac{C_{\alpha\beta\mu\nu}C^{\alpha\beta\mu\nu}}{R_{\mu\nu}R^{\mu\nu}}$, where $P^2$ represents the gravitational entropy and vanishes at the initial singularity of the universe. This is called Penrose’s Weyl Curvature Conjecture, according to which
the Ricci part of the curvature dominates over the Weyl part at the initial singularity of the universe.
 
GR admits solutions that include essential, physical singularities which are regions where some scalar curvature diverges at a point in question. Physical singularities indicate a breakdown of
GR, but it is expected that a quantum theory of gravity will somehow cure this. Investigators have been examining proposed corrections to GR that bound certain curvature invariants to values less than appropriate powers of the Planck length within the context of cosmologies and low-dimensional black holes. Loop Quantum Gravity (LQG) uses curvature invariants to probe
physical singularities in homogeneous spacetimes, black hole spacetimes, and inhomogeneous Gowdy spacetimes \cite{Singh}. There is confirmation from phenomenological, analytical,
and numerical investigations that singularities are resolved for various homogeneous spacetimes in Loop Quantum Cosmology (LQC) wherein cosmological spacetimes are quantized using the methods of LQG \cite{Singh}. There’s also evidence for singularity resolution
for the various isotropic and anisotropic Bianchi spacetimes using the LQG methods. Singularities are resolved via non-perturbative gravitational effects because of LQG’s discrete
quantum geometry which bounds the curvature invariants by the Planck curvature scale. This leads to the replacement of the big bang by a big bounce close to the Planck scale which is a
robust result in LQC \cite{Singh}.


Recent developments in computer algebra systems \cite{MacCallum:2018csx} have provided researchers with opportunities to perform lengthy or complex calculations more easily than ever before. The calculation of the  ZM or CM curvature invariants for a given metric are examples of such lengthy and time consuming calculation, and this is now achievable on most desk-top computers. As we have emphasized throughout this essay, calculation of these invariants is beneficial for many areas of theoretical physics such as GR, black hole thermodynamics and quantum gravity. It is worth mentioning that the calculation and analysis of curvature invariants can still present research opportunities for talented graduate students who are interested in the interface between theoretical physics and computer programming.


\end{document}